\documentclass{emulateapj}
\usepackage{natbib}
\usepackage{graphicx}
\usepackage{epstopdf}
\usepackage{color}

\newcommand{\kms}{km s$^{-1}$} 
\newcommand{\Ha}{H$\alpha$}
\newcommand{\Hb}{H$\beta$}
\newcommand{\oiii}{[OIII]$\lambda$5007 }

\newcommand{\LyA}{Ly$\alpha$}

\newcommand{\fesc}{$f^{Ly\alpha}_{esc}$}

\begin{document}
\title{Green Pea Galaxies Reveal Secrets of \LyA\ Escape}

\author{Huan Yang\altaffilmark{1,2}, Sangeeta Malhotra\altaffilmark{2},
Max Gronke\altaffilmark{3}, James E. Rhoads\altaffilmark{2}, Mark Dijkstra\altaffilmark{3}, 
Anne Jaskot\altaffilmark{4}, Zhenya Zheng\altaffilmark{5,6}, and Junxian Wang\altaffilmark{1} }

\altaffiltext{1}{CAS Key Laboratory for Research in Galaxies and Cosmology, Department of Astronomy, University of Science and Technology of China; yanghuan@mail.ustc.edu.cn}
\altaffiltext{2}{Arizona State University, School of Earth and Space Exploration; huan.y@asu.edu; Sangeeta.Malhotra@asu.edu; James.Rhoads@asu.edu}
\altaffiltext{3}{Institute of Theoretical Astrophysics, University of Oslo, Norway}
\altaffiltext{4}{Smith College, Northampton, MA}
\altaffiltext{5}{Pontificia Universidad Cat\'{o}lica de Chile, Santiago, Chile}
\altaffiltext{6}{Chinese Academy of Sciences South America Center for Astronomy, Santiago, Chile}

\begin{abstract}

We analyze archival \LyA\ spectra of 12 ``Green Pea" galaxies observed with the Hubble Space Telescope, model their \LyA\ profiles with radiative transfer models, and explore the dependence of \LyA\ escape fraction on various properties. Green Pea galaxies are nearby compact starburst galaxies with [OIII]$\lambda$5007 equivalent widths of hundreds of \AA. All 12 Green Pea galaxies in our sample show \LyA\ lines in emission, with a  \LyA\ equivalent width distribution similar to  high redshift \LyA\ emitters. Combining the optical and UV spectra of Green Pea galaxies, we estimate their \LyA\ escape fractions and find correlations between \LyA\ escape fraction and kinematic features of \LyA\ profiles. The escape fraction of \LyA\ in these galaxies ranges from 1.4\% to 67\%. We also find that the \LyA\ escape fraction depends strongly on metallicity and moderately on dust extinction.  We compare their high-quality \LyA\ profiles with single HI shell radiative transfer models and find that the \LyA\ escape fraction anti-correlates with the derived HI column densities. Single shell models fit most \LyA\ profiles well, but not the ones with highest escape fractions of \LyA.  Our results suggest that low HI column density and low metallicity are essential for \LyA\ escape, and make a galaxy a \LyA\ emitter.  
\end{abstract}

\section{Introduction}

The \LyA\ emission line is a key tool in  discovering and studying high redshift galaxies, and a good probe for reionization. High redshift \LyA\ emission line galaxies (LAE) have been found routinely for almost two decades (e.g. Dey et al. 1998; Hu et al. 1998; Rhoads et al. 2000; Ouchi et al. 2003; Gawiser et al. 2006; Guaita et al. 2010; Cl{\'e}ment et al. 2012; Shibuya et al. 2012; Matthee et al. 2014). 
These high redshift LAEs are generally small, young star forming galaxies. They have compact size, low stellar mass, low dust extinction, low metallicity, and high specific star formation rate (sSFR) (e.g. Malhotra 2012; Bond et al. 2010; Gawiser et al. 2007; Pirzkal et al. 2007; Finkelstein et al. 2008; Pentericci et al. 2009). 
At $2\lesssim z \lesssim 6$, these LAEs are an important population of star-forming galaxies, and they constitute an increasing fraction of Lyman break galaxies across that range, reaching $\sim$60\% of Lyman break galaxies (LBGs) at redshift z$\sim$6 (Stark et al. 2011). At $z>6$, the \LyA\ luminosity, \LyA\ equivalent width (EW), and spatial clustering of LAEs and LBGs are used to probe reionization of the intergalactic medium (e.g. Malhotra \& Rhoads 2004; Ouchi et al. 2010; Hu et al. 2010; Kashikawa et al. 2011; Ota et al. 2010; Treu et al. 2012; Pentericci et al. 2014; Tilvi et al. 2014). 

To use \LyA\ as a powerful probe of high redshift galaxies and reionization, we should ideally understand how \LyA\ escapes from galaxies (e.g. Dijkstra et al. 2014). In star-forming galaxies, \LyA\ photons come from the recombination of gas surrounding hot young O and B stars. Because \LyA\ is resonantly scattered, the path length and dust extinction of \LyA\ photons is increased in a manner that depends on the kinematics of the gas.  This process also determines the characteristic profile of \LyA\ lines. Only a fraction of the intrinsic \LyA\ photons can escape the galaxy and be observed. The emergent \LyA\ emission depends on the amount of dust, the HI gas column density ($N_{HI}$), the velocity distribution of HI gas, and the geometric distribution of HI gas and dust (e.g. Neufeld 1990; Charlot \& Fall 1993; Ahn et al. 2001; Verhamme et al. 2006; Dijkstra et al. 2006).

Since the scattering of \LyA\ photons can significantly modify the \LyA\ profile, studying \LyA\ profiles is an important way to understand \LyA\ escape. \LyA\ emission line in LAEs usually shows an asymmetric or a double-peaked profile (e.g. Rhoads et al 2003; Kashikawa et al. 2011; Erb et al. 2014).   
For LAEs with detected optical emission lines and systemic redshifts, the peaks of \LyA\ profiles are usually redshifted with respect to systemic velocities (McLinden et al. 2011, 2014; Chonis et al. 2013; Hashimoto et al. 2013; Song et al. 2014; Shibuya et al. 2014; Erb et al. 2014). 
The velocity offset of \LyA\ emission line from systemic velocity is usually smaller in LAEs than in continuum selected galaxies with weaker \LyA\ emission lines or \LyA\ absorption (Shapley et al. 2003).

To understand \LyA\ profile, the single shell outflow model is popular for its simplicity and ability to capture several essential features of \LyA\ lines. In a galaxy with spherical HI gas outflow, \LyA\ photons backscattered from the far side of the receding shell will acquire a frequency shift that allows them to pass through other gas in the galaxy, including the near side of the shell. 
Many studies simulate the \LyA\ radiative transfer process assuming a single such shell (e.g. Ahn et al. 2001; Verhamme et al. 2006; Dijkstra et al. 2006; Schaerer et al. 2011; Gronke et al. 2015). 
The output \LyA\ profile depends on the HI column density of the gas shell, the velocity of the shell, the dust optical depth, and the temperature of the HI gas in the shell. 
The models can reproduce the observed \LyA\ profile and infer properties such as  $N_{HI}$ and outflow velocity of the HI shell.  In these models, the \LyA\ velocity offset is formed from scattering of \LyA\ photons by gas outflows, and increasing $N_{HI}$ or outflow velocity will usually result in larger \LyA\ velocity offset.

To get a better understanding of \LyA\ escape, we need to observe high quality \LyA\ profiles, and determine systemic redshifts, gas outflows, the HI gas distribution/kinematics, and many other galactic properties of LAEs.  At high redshift, however, absorption by the intergalactic \LyA\ forest prevents reliable measurements 
of the blue portion of \LyA\ emission lines.  Other crucial observations are also impractical, both because high-$z$ LAEs are faint, and because some features (notably rest-optical emission lines) are redshifted to $\lambda_{obs} > 2.4{\mu}m$, where presently available instruments lack sensitivity.
Therefore many studies seek to solve the \LyA\ escape problem in the nearby universe by observing galaxies with similar properties to high-z LAEs (e.g. Giavalisco et al. 1996; Kunth et al. 1998; Mas-Hesse et al. 2003; Deharveng et al. 2008; Finkelstein et al. 2009; Atek et al. 2009; Scarlata et al. 2009; Leitherer et al. 2011; Heckman et al. 2011; Cowie et al. 2011; Wofford et al. 2013; Hayes et al. 2005, 2014; Ostlin et al. 2009, 2014; Pardy et al. 2014; Rivera-Thorsen et al. 2015). 
However, these samples of nearby \LyA\ emission line galaxies have on average much smaller \LyA\ equivalent widths and \LyA\ escape fractions (1\%-12\%  in Wofford et al. (2013) sample; Hayes et al. 2014) than do high-z LAE samples (\textgreater17\%  Zheng et al. 2012; Nakajima et al. 2012).  The main exception is the present Green Pea galaxy sample.

Green Pea galaxies were discovered in the citizen science project Galaxy Zoo, in which public volunteers morphologically classified millions of galaxies from the Sloan Digital Sky Survey (SDSS). 
These are compact galaxies that are unresolved in SDSS images. The green color is because the [OIII] doublet dominates the flux of SDSS $r$-band which is mapped to the green channel in the SDSS's false-color {\it gri}-band images. 
They have redshifts $0.11 < z < 0.36$, small stellar masses $\sim10^{8}-10^{10} M_{\odot}$, low metallicities, high specific star formation rates (sSFR), and emission line equivalent widths (EW(\Ha) and EW(\oiii) exceeding hundreds of \AA\ (Cardamone et al. 2009; Izotov et al. 2011).  Thus Green Peas are good counterparts to high-redshift LAEs in size, morphology, stellar mass, metallicities, and optical emission line strengths.  This suggested that Green Peas might also show strong \LyA\ emission lines. 

In this paper, we use archival HST \LyA\ spectroscopy of Green Peas to show that Green Peas are the best analogs of high-z LAEs in local universe. With high quality \LyA\ spectra and rest-frame optical spectra, Green Peas provide a good opportunity to study \LyA\ escape. We explore relations of \LyA\ escape to \LyA\ profiles and to galactic properties, compare the \LyA\ profiles with radiative transfer models, and discuss constrains on HI gas and \LyA\ escape.

\section{Green Peas Sample and Spectra Data}
\subsection{Green Peas Sample}
While the full SDSS data set contains a few tens of thousands of Green Peas candidates based on photometric selection, they are not in a category systematically targeted for spectroscopic followup.
In SDSS DR7, a sample of about 251 Green Peas were observed as serendipitous spectroscopic targets (Cardamone et al. 2009).   Which Green Peas were thus targeted was essentially a random process,
depending where the SDSS fibers were undersubscribed by other objects of all types.  Thus, we do not expect this step of the selection to introduce any important bias in the sample.
A subset of these objects have sufficient signal to noise ratio (S/N) in both continuum and emission lines (\Ha, \Hb,  and \oiii) to study galactic properties such as SFR, stellar mass, and metallicity.  Galaxies with an active galaxies nucleus (AGN) (diagnosed by their broad Balmer emission lines or \Ha/[NII] vs. [OIII]/\Hb\ diagram) are excluded. These selections result in 66 Green Peas that have good optical spectra and measured galactic properties (see Cardamone et al. 2009 and Izotov et al. 2011 for details about selection of Green Peas sample). This provides a parent sample of Green Peas for \LyA\ emission studies. 

We searched for UV spectra of the parent sample of 66 Green Peas in the HST archive, and find 12 Green Peas that have UV and \LyA\ spectra taken with the Cosmic Origins Spectrograph (COS) (PIs: Henry (GO: 12928); Jaskot (GO: 13293); Heckman (GO: 11727)). We study this sample of 12 Green Peas in this paper.  The 9 galaxies in Henry's sample were selected from the parent sample of 66 Green Peas by their FUV brightness, with $m_{FUV}<20\ (AB)$.  The 2 galaxies in Jaskot's sample were selected by their extreme [OIII]/[OII] ratios. 
The one galaxy in Heckman's sample was selected by its high FUV luminosity, high UV flux, and compact size. 
Compared to the parent sample, the current sample of 12 Green Peas covers the full ranges of mass, SFR, and EW([OIII]) of the parent sample, and these 12 Green Peas are only slightly biased to lower metallicity and lower dust extinction (see figure 1). 
To address the bias of the current 12 Green Peas sample and explore the relations between \LyA\ and galactic properties, we are taking \LyA\ spectra of a larger sample of Green Peas to cover more completely the distributions of metallicities and dust extinctions (PI: Malhotra (GO: 14201)).

\begin{figure}[]
\centering
  \includegraphics[width=0.5\textwidth]{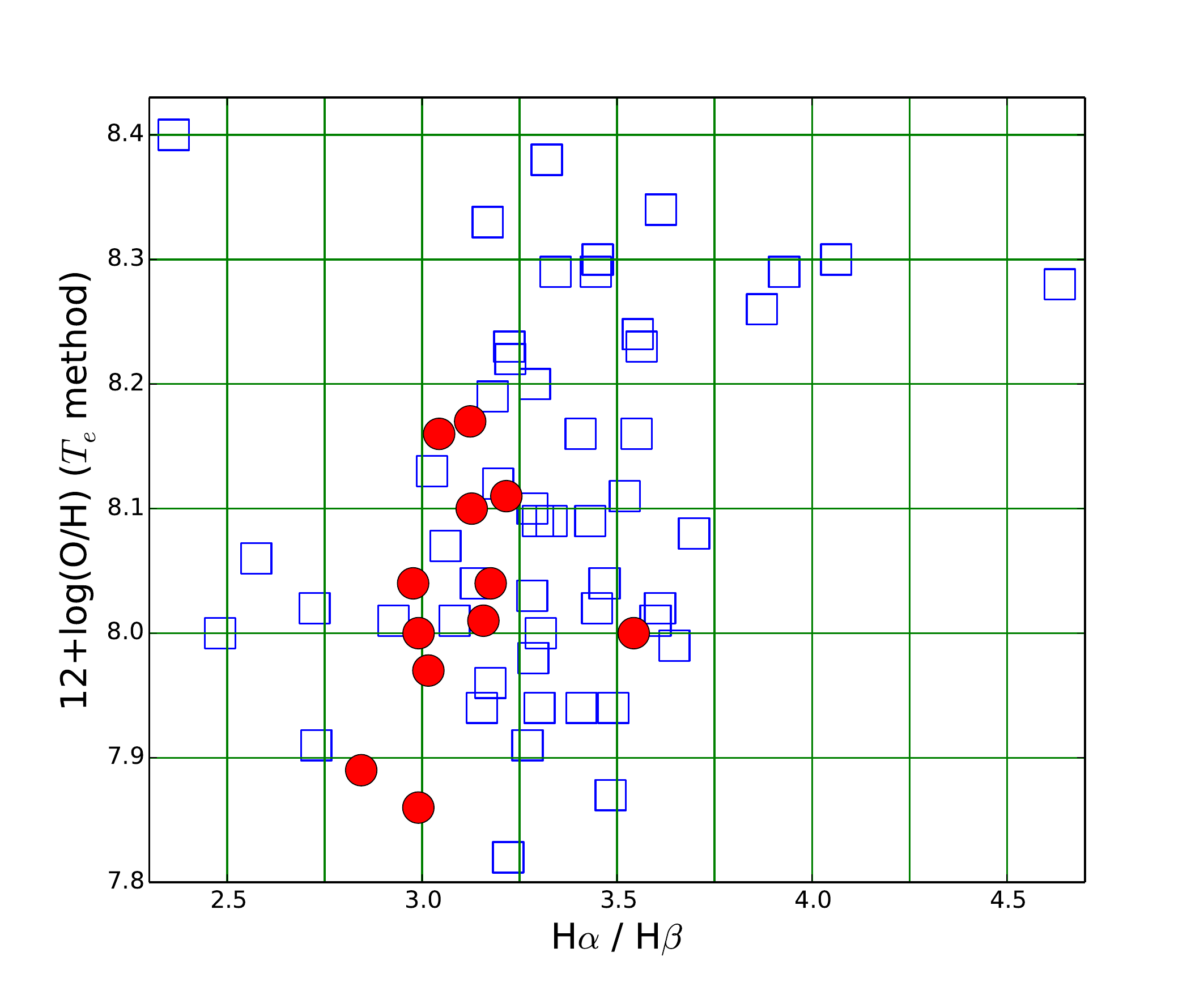}
 \caption{Green Pea sample distribution in the diagram of \Ha/\Hb\ and metallicity. \Ha/\Hb\ is the flux ratio of \Ha\ and \Hb\ emission lines.  In the absence of dust reddening, we expect \Ha/\Hb $\approx 2.86$, and
higher values of the ratio indicate the presence of dust.  Red dots are the current sample of 12 Green Peas with \LyA\ spectra. Blue empty squares are the parent sample of 66 Green Peas. }
\end{figure}

\begin{deluxetable*}{lllccccccccc}[!ht]
\tabletypesize{\scriptsize}
\centering
\tablewidth{0pt}
\tablecaption{Sample}

\tablehead{\colhead{ID} & \colhead{RA} & \colhead{DEC} & \colhead{Redshift} &  \colhead{E(B-V)$_{MW}$}  &  \colhead{\LyA\ flux} & \colhead{EW(\LyA)} & \colhead{\fesc} & \colhead{12+log(O/H)} & \colhead{E(B-V)}  \\ 
\colhead{} & \colhead{J2000} & \colhead{J2000} & \colhead{} & \colhead{mag}  & \colhead{$10^{-14}$ erg~s$^{-1}$~cm$^{-2}$~} & \colhead{\AA} & \colhead{} & \colhead{} & \colhead{mag} \\
\colhead{(1)} & \colhead{(2)} & \colhead{(3)} & \colhead{(4)} & \colhead{(5)} & \colhead{(6)} & \colhead{(7)} & \colhead{(8)} &\colhead{(9)} &\colhead{(10)}  }
\startdata
GP1457+2232 & 14:57:35.13 & +22:32:01.8 & 0.148611 & 0.0410 & 0.46 & 7.36 & 0.014 & 8.04 & 0.061 \\
GP0303-0759 & 03:03:21.41 & -07:59:23.2 & 0.164880 & 0.0845 & 1.01 & 7.20 & 0.050 & 7.86 & 0.000 \\
GP1244+0216 & 12:44:23.37 & +02:15:40.5 & 0.239426 & 0.0211 & 1.89 & 39.96 & 0.065 & 8.17 & 0.062 \\
GP1054+5238 & 10:53:30.83 & +52:37:52.9 & 0.252638 & 0.0126 & 1.54 & 10.66 & 0.068 & 8.10 & 0.069 \\
GP1137+3524 & 11:37:22.14 & +35:24:26.7 & 0.194390 & 0.0156 & 3.81 & 33.43 & 0.130 & 8.16 & 0.043 \\
GP0911+1831 & 09:11:13.34 & +18:31:08.2 & 0.262200 & 0.0243 & 3.15 & 49.53 & 0.155 & 8.00 & 0.168 \\
GP0926+4428 & 09:26:0.44 & +44:27:36.5 & 0.180690 & 0.0156 & 6.36 & 40.82 & 0.245 & 8.01 & 0.074 \\
GP1424+4217 & 14:24:05.73 & +42:16:46.3 & 0.184788  & 0.0087 & 8.55 & 78.27 & 0.266 & 8.04 & 0.028 \\
GP0815+2156 & 08:15:52.00 & +21:56:23.6 & 0.140950 & 0.0352 & 4.01 & 75.09 & 0.299 & 8.00 & 0.014 \\
GP1133+6514 & 11:33:03.80 & +65:13:41.3 & 0.241397 & 0.0093 & 2.08 & 35.29 & 0.352 & 7.97 & 0.040 \\
GP1249+1234 & 12:48:34.64 & +12:34:02.9 & 0.263389 & 0.0256 & 5.28 & 94.80 & 0.383 & 8.11 & 0.084 \\
GP1219+1526 & 12:19:03.98 & +15:26:08.5 & 0.195599 & 0.0224 & 13.45 & 157.54 & 0.672 & 7.89 & 0.000 
\enddata
\tablecomments{Column Descriptions: (1) Object ID; (4) The Milky Way extinction $E(B-V)_{MW}$, based on Schlafly \& Finkbeiner (2011); (6) \LyA\ flux measured from the spectra, and corrected for Milky Way dust extinction using the Fitzpatrick (1999) extinction law; (7) rest-frame \LyA\ equivalent width; (8) \LyA\ escape fraction;  (9) metallicity measured from direct $T_{e}$ method by Izotov et al. 2011; (10) Dust extinction of Green Pea galaxies. 
Errors of these measurements are dominated by systematics, so statistical errors are not given. }
\end{deluxetable*}

\begin{deluxetable*}{lccccccc}

\tablecaption{\LyA\ Profile Analysis}

\tablehead{\colhead{ID} & \colhead{V(blue-peak)} & \colhead{V(red-peak)} & \colhead{V(valley)}& \colhead{FWHM(\Ha)} & \colhead{FWHM(red)} & \colhead{${Flux(blue)}\over{Flux(red)}$} & \colhead{${f_{\lambda}(valley)}\over{f_{\lambda}(red-peak)}$} \\ 
\colhead{} & \colhead{\kms} & \colhead{\kms} & \colhead{\kms} & \colhead{\kms} & \colhead{\kms} & \colhead{} & \colhead{}  \\
\colhead{(1)} & \colhead{(2)} & \colhead{(3)} & \colhead{(4)} & \colhead{(5)} & \colhead{(6)} & \colhead{(7)} & \colhead{(8)}} 

\startdata
GP1457+2232 & -329 & 406 & 65 & 188 & 350 & 0.410 & 0.013 \\
GP0303-0759 & -313 & 153 & -106 & 233 & 273 & 0.047 & 0.023 \\
GP1244+0216 & -240 & 247 & -21 & 233 & 268 & 0.348 & 0.005 \\
GP1054+5238 & -266 & 192 & -1 & 291 & 230 & 0.108 & 0.015 \\
GP1137+3524 & -355 & 201 & -27 & 287 & 253 & 0.120 & 0.022 \\
GP0911+1831 & -278 & 81 & -62 & 298 & 224 & 0.185 & 0.033 \\
GP0926+4428 & -223 & 244 & -38 & 280 & 307 & 0.154 & 0.071 \\
GP1424+4217 & -150 & 224 & 78 & 259 & 187 & 0.631 & 0.059 \\
GP0815+2156 & -121 & 144 & -15 & 193 & 185 & 0.432 & 0.115 \\
GP1133+6514 & -69 & 271 & 125 & 197 & 219 & 0.399 & 0.266 \\
GP1249+1234 & -$^{a}$ & 83 & -$^{a}$ & 209 & 322 & -$^{a}$ & -$^{a}$ \\
GP1219+1526 & -76 & 176 & 25 & 244 & 177 & 0.389 & 0.114 
\enddata
\tablecomments{Column Descriptions: (1) Object ID;  (2) V(blue-peak) is the velocity of the peak in the blue side of \LyA\ profile; (3) V(valley) is the velocity of the center valley of \LyA\ profile;
(4) V(red-peak) is the velocity of the peak in the red side of \LyA\ profile; (5) FWHM of \Ha\ emission line from SDSS spectra; (6) FWHM(red) is the FWHM of \LyA\ red peak; (7) Flux(blue)/Flux(red) is the flux ratio of blue and red part of \LyA\ profile;  (8) $f_{\lambda}(valley)/f_{\lambda}(red-peak)$ is the flux density ratio of central valley to red peak of \LyA\ profile.  All velocities are relative to the peak of the \Ha\ emission.}
\tablenotetext{a}{ GP1249+1234 doesn't show double-peaked \LyA\ profile. }

\end{deluxetable*}

\begin{figure*}[ht]
\centering
  \includegraphics[width=\textwidth]{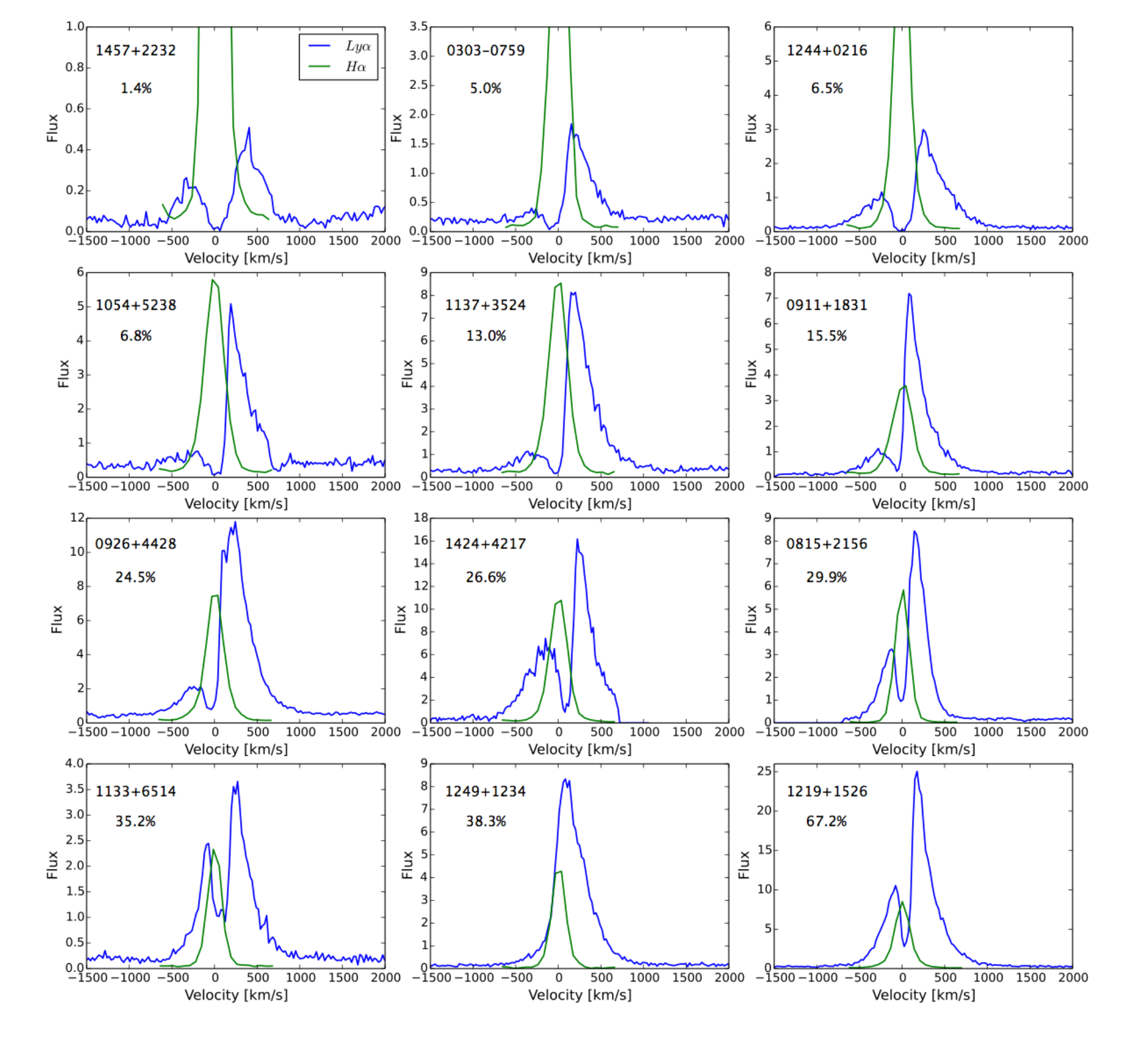}
 \caption{ \LyA\ and \Ha\ emission line profiles of Green Peas. \LyA\ and \Ha\ are in the same flux units of $10^{-17} erg\ cm^{-2}\ s^{-1}\ (km\ s^{-1})^{-1}$ for all plots. These 12 galaxies are sorted by increasing \fesc\ from left to right, and top to bottom. The \fesc\ is given in each panel.}
\end{figure*}

\begin{figure}[h]
\centering
  \includegraphics[width=0.5\textwidth]{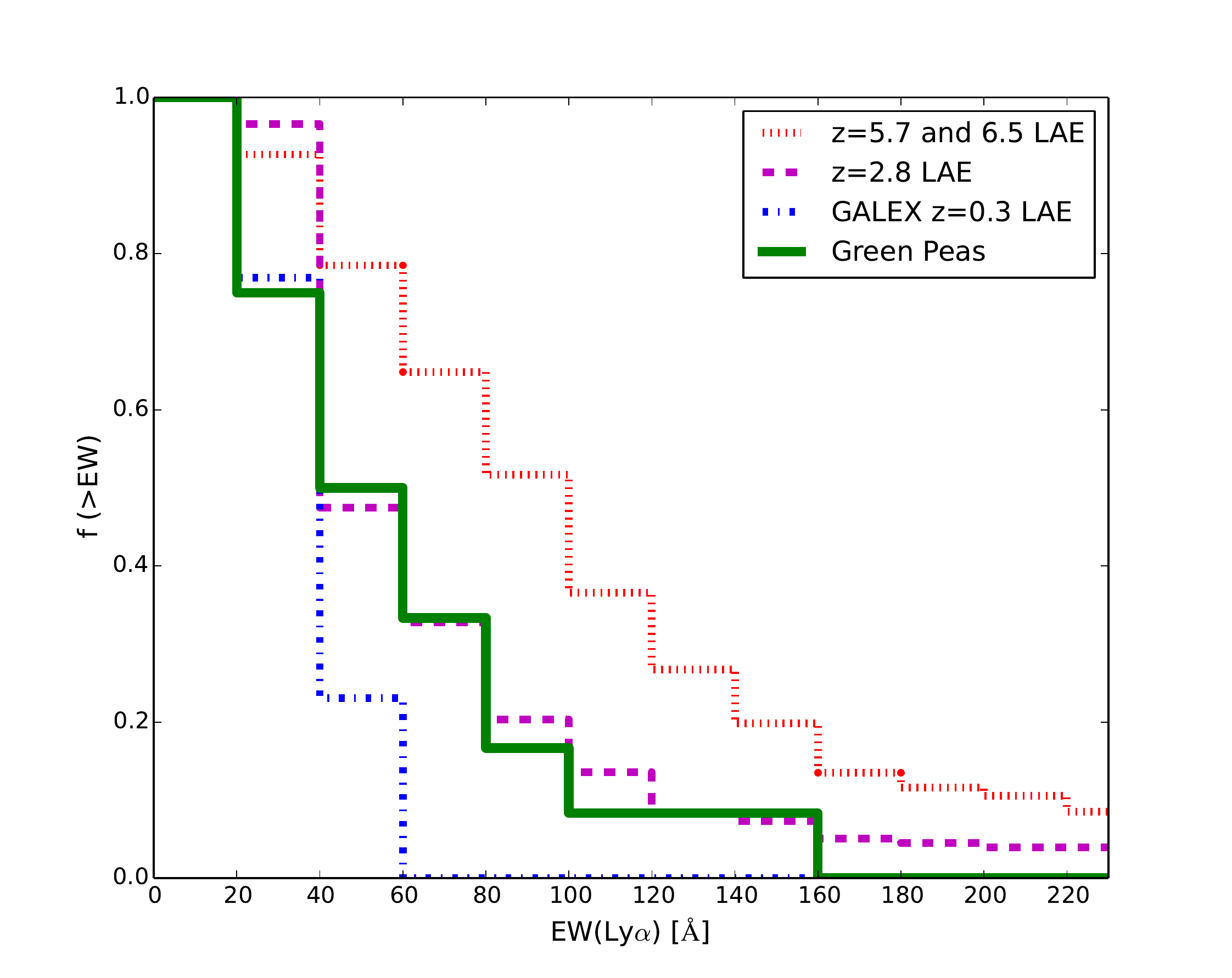}
 \caption{Here we compare the rest-frame EW(\LyA) distribution of Green Peas with different samples. The solid green line shows the 12 Green Peas. The blue dash-dot line shows the GALEX z=0.3 LAE sample (Cowie et al. 2011; Finkelstein et al.  2009; Scarlata et al. 2009).  The dashed magenta line shows the z=2.8 LAE sample from Zheng et al. (2016, in preparation). The dotted red line shows the z=5.7 and 6.5 LAE sample from Kashikawa et al. (2011). The EW(\LyA) distribution of Green Peas is similar to the z=2.8 LAE sample. }
\end{figure}

\subsection{UV and Optical Spectra}
12 Green Peas were observed by HST-COS. All were imaged using the COS acquisition modes ACS/IMAGE, which took high resolution NUV acquisition images and centered the target accurately (error $\sim$ 0.05 arc-second) in the 2.5 arc-seconds diameter Primary Science Aperture. These images show the NUV sizes of Green Peas are compact compared to the 2.5 arc-seconds aperture.   
Rest-frame spectral coverages are roughly $\sim$ 950-1500\AA. \LyA\ spectra were taken with both FUV grating G130M and G160M for two sources. Since the \LyA\ were detected at very high S/N ratio, we only use G160M spectra for these two sources. 9 sources have \LyA\ spectra taken only with G160M. One source has \LyA\ spectra taken only with G130M. 

We retrieved COS spectra for these 12 Green Peas from the HST MAST archive after they have been processed through the standard COS pipeline CALCOS version 3.0 (2014-10-30). The resulting spectral resolutions are coarser than the point-source spectral resolution (FWHM$\sim$20 \kms), because these Green Peas are resolved in their NUV acquisition images. Their real spectral resolution depends on source angular sizes. For FUV continuum the resolutions are about 20-50 \kms\ based on their NUV sizes (James et al. 2014; Henry et al. 2015). For \LyA\ emission line spectra, the resolution may be somewhat worse if \LyA\ is more extended than the UV continuum emission. If \LyA\ filled the aperture uniformly, the spectral resolution would be $\sim$200\ \kms\ FWHM (France et al. 2009). 
We bin the reduced COS spectra to $\sim$ 0.12\AA\ $pixel^{-1}$ for 11 spectra taken with G160M and to $\sim$ 0.08\AA\ $pixel^{-1}$ for the one spectra taken with G130M. The velocity precisions of the UV spectra are better than 40 \kms\ (Henry et al. 2015). The precisions of the systemic redshifts from \Ha\ emission lines in SDSS spectra are better than 20 \kms. The resulting \LyA\ line profiles are shown in figure~2. 

Then we measure the properties of \LyA\ emission lines.  In most cases, we estimate the continuum level from rest-frame wavelength range $\sim$ 1225-1260\AA. For GP1457+2232 (see Table 1 for the source ID), the \LyA\ spectrum shows damped absorption wings, and we measure the continuum from rest-frame wavelength range $\sim$ 1270-1300\AA. For GP1137+3524, we don't have spectra red-ward of \LyA, so we measure the continuum from rest-frame wavelength range $\sim$ 1180-1210\AA. To get the \LyA\ line flux, we integrate the continuum subtracted spectra in rest-frame wavelength range $\sim$ 1210-1220\AA. For GP1457+2232 with damped absorption, we get its residual \LyA\ emission line flux in the center of damped absorption without subtracting the continuum. We correct for Milky Way extinction using the attenuation determined by Schlafly \& Finkbeiner (2011) and the Fitzpatrick (1999) extinction law. The foreground extinction values were obtained from the NASA/IPAC Galactic Dust Reddening and Extinction tool. 

We downloaded optical spectra of these Green Peas from the SDSS DR12 archive,  along with the pipeline measurements of their emission line properties. We correct the measured \Ha\ and \Hb\ fluxes for Milky Way extinction (again using the Fitzpatrick (1999) extinction law). Then we calculate E(B-V) from dust in Green Peas galaxies using Calzetti et al. (2000) extinction law and an intrinsic \Ha/\Hb\ ratio of 2.86, and correct dust reddening of the observed \Ha\ flux. We use the metallicities and mass measured from SDSS spectra by Izotov et al. (2011). The metallicities were calculated using $T_{e}$ method. These properties of Green Pea galaxies and their \LyA\ emission lines are shown in Table~1. 

Then we calculate the rest-frame EW(\LyA), EW(\LyA)=Flux(\LyA)/$f_{\lambda}(continuum)$/(1+redshift), and the \LyA\ escape fraction \fesc.  \fesc\ is defined as the ratio of observed \LyA\ flux to intrinsic \LyA\ flux. Assuming case-B recombination, the intrinsic \LyA\ flux is 8.7 times dust extinction corrected \Ha\ flux. Thus the \fesc\ is \LyA/(8.7$\times H\alpha_{corrected}$).  The SDSS \Ha\ spectra were taken with 3 arc-seconds diameter aperture which matches the COS 2.5 arc-seconds diameter aperture very well.

\section{Green Peas Are Analogs of High-z LAEs} 
{\it The first remarkable result is all 12 Green Peas show \LyA\ emission lines} (figure 2; and see also Henry et al. (2015) for 10 of these 12 Green Peas). Furthermore, 9 of 12 Green Peas have EW(\LyA) larger than 30 \AA, and would be selected as LAEs in high redshift samples. This is in contrast to other low-redshift samples of star-forming galaxies and \LyA\ galaxies. 
The LARS sample (Ostlin et al. 2014), selected on the basis of high EW(\Ha), shows \LyA\ in emission in 12 of 14 galaxies, but only 6/14 would be picked up as high-z \LyA\ emitters on the basis of EW(\LyA) alone (Hayes et al. 2014). Kunth et al. (1998) detected \LyA\ emission in 4 of 8 low-$z$ galaxies in their sample, with a mean EW(\LyA) of 26 \AA\ among the detections.  Wofford et al. (2013) detected \LyA\ emission in 7 of 20 low-z star-forming galaxies, and all detections have EW(\LyA)$<$12 \AA.  The GALEX z$\sim$0.3 \LyA\ emitter sample, selected with EW(\LyA)$\gtrsim$15\AA, has mean EW(\LyA) of 30 \AA\ (Deharveng et al. 2008; Cowie et al. 2011; Finkelstein et al.  2009; Scarlata et al. 2009).

We then compare the EW(\LyA) distribution of Green Peas with nearby and high-z LAE samples (figure 3).  For the high-z samples, we use a $z=2.8$ narrow-band selected LAE sample (Zheng et al. in preparation) and a sample of spectroscopically confirmed LAEs at z=5.7 and 6.5 (Kashikawa et al. 2011). All known AGNs in these high-$z$ samples are excluded, and the AGN contamination for high-$z$ LAE samples is less than 5\% (Zheng et al. 2013). The EW(\LyA) of high-z samples are computed from deep narrow-band and broad-band photometries. 
In figure~3, we compare the cumulative normalized EW(\LyA) distribution for these samples. The EW(\LyA) distribution of Green Peas is very similar to the high redshift ($z=2.8$) sample.
According to K-S test results, the EW(\LyA) distribution of the GALEX LAE sample differs from that of the $z=2.8$ LAE sample at a probability of 99\%, while the EW(\LyA) distribution of Green Peas sample is not distinguishable from that of the $z=2.8$ LAE sample .

Of the two Green Peas (GP1457+2232 and GP0815+2156) selected by extreme [OIII]/[OII] ratio, 
one shows weak \LyA\  and the other shows strong \LyA\ emission (figure 2). They are not especially distinguishable in their \fesc\ or EW(\LyA) from the other Green Peas. And we don't see correlations between \fesc\ and [OIII]/[OII] ratio in this sample of 12 Green Peas. So including the two sources with extreme [OIII]/[OII] ratio doesn't bring obvious bias to this sample.

\begin{figure*}[ht]
\centering
  \includegraphics[width=\textwidth]{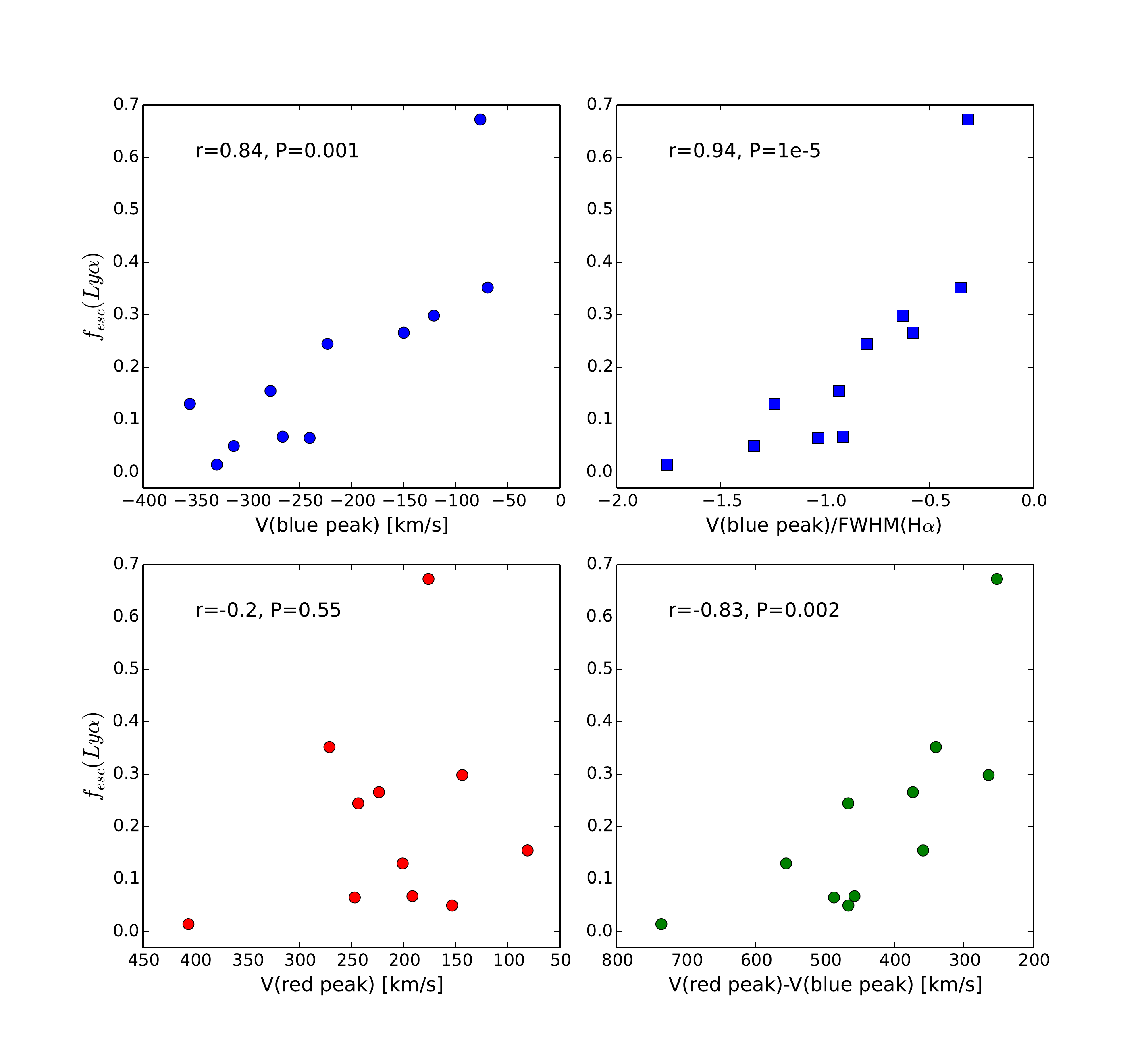}
 \caption{Relations between \fesc\ and velocity quantities of \LyA\ profile. (Upper-left): \fesc\ and velocity of blue peak of \LyA\ profile. (Upper-right): \fesc\ and velocity of blue peak normalized by FWHM(\Ha). (Lower-left): \fesc\ and velocity of red peak of \LyA\ profile. (Lower-right): \fesc\ and peak seperation of \LyA\ profile. The Spearman correlation coefficient r, and null probability P, are given in each panel. }
\end{figure*}

\begin{figure*}[h]
\centering
  \includegraphics[width=\textwidth]{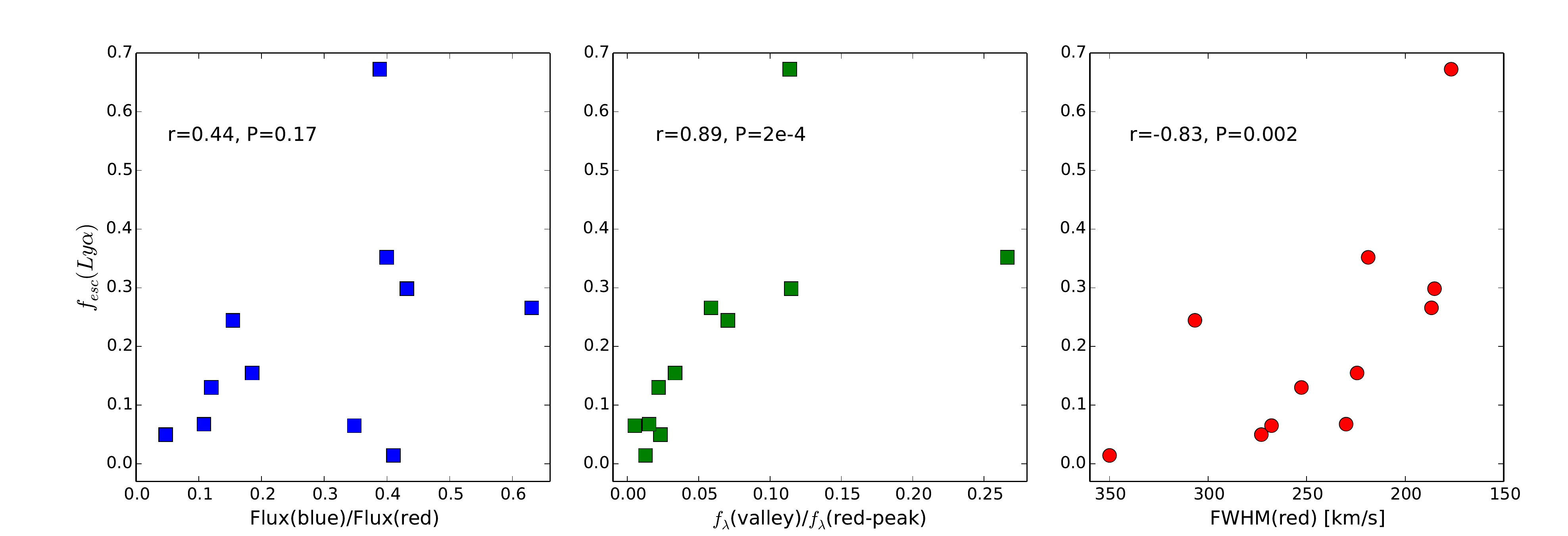}
 \caption{Relations between \fesc\ and:  (Left):  the flux ratio of the blue to the  red part of the \LyA\ profile. (Middle): the flux density ratio of the valley to the red peak of the \LyA\ profile. (Right): the FWHM of the red part of the \LyA\ profile. The Spearman correlation coefficient r, and null probability P, are given in each panel.}
\end{figure*}

\begin{figure*}[h]
\centering
  \includegraphics[width=\textwidth]{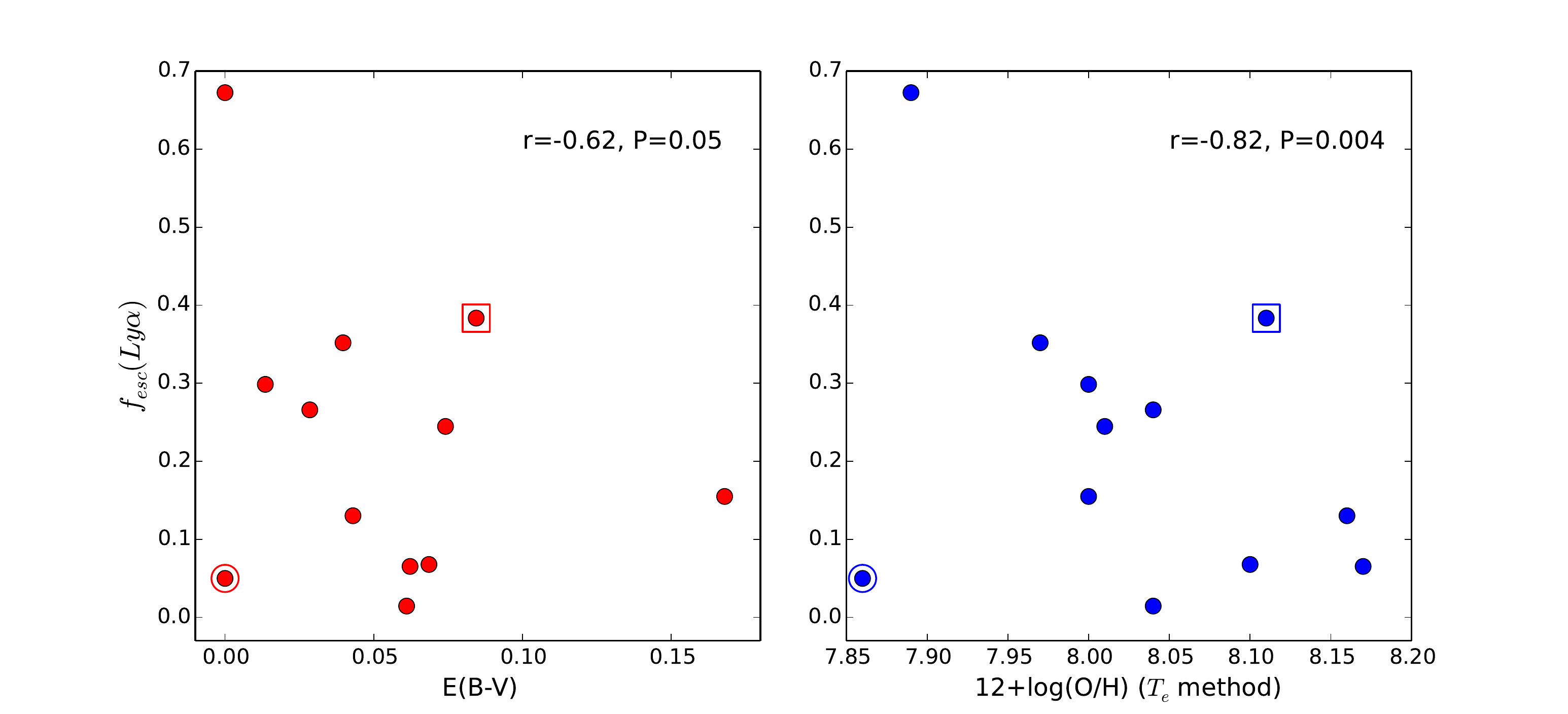}
 \caption{Relations between \fesc\ and: (Left): dust extinction E(B-V). (Right): metallicity. The Spearman correlation coefficient r, and null probability P, are given in each panel. GP0303-0759 (marked by a large circle) has the lowest metallicity and E(B-V). Its WISE three band colors suggest it is probably an AGN (Malhotra et al. in preparation). GP1249+1234 (marked by a large square) has single peak profile and large E(B-V). We exclude both sources when calculating correlation coefficients.}
\end{figure*}

\section{\LyA\ Profiles Analysis \& \LyA\ Escape}
We now use Green Peas to study the mechanisms of \LyA\ escape. 
The escape of the resonantly scattered \LyA\ photons involves potentially complex interactions with gas and dust.  This results in richly varied \LyA\ line profiles that carry a great deal of information, especially when the profiles are double-peaked as they tend to be for the Green Peas.  We therefore begin by exploring relations between \LyA\ escape fraction and \LyA\ profiles.

In figure 2, we display the \LyA\ profiles of these 12 Green Peas in order of increasing \LyA\ escape fraction from left to right, and top to bottom.  The zero point of the velocity scale is set by the peak of the H$\alpha$ 
line.    We define the ``red peak'' as the peak in the \LyA\ line profile occurring at velocity \textgreater\ 0. 
11 of the 12 galaxies have double-peaked line profiles, and for these we define the ``blue peak'' as the \LyA\ peak at velocity \textless\ 0,  and the ``valley'' as the flux minimum between the two peaks for double peak profiles.  Simply looking at their profiles, we can find that as \fesc\ increases, the residual flux at the inter-peak valley gets stronger, and the blue peak moves nearer to the systemic velocity.  To characterize those trends quantitatively, we measure the velocity and flux density $f_{\lambda}$ at the red peak, blue peak, and valley. We also measure the integrated flux and full width at half maximum (FWHM) for the red (V\textgreater0) and blue (V\textless0) portions of the \LyA\  profile -- Flux(red), Flux(blue), FWHM(red), and FWHM(blue).  Then we investigate correlations between those quantities and \LyA\ escape fraction.

As shown in figure 4, \fesc\ strongly correlates with the blue peak velocity V(blue-peak) (see also Henry et al. 2015) for the 11 Green Peas with double peaked profiles. The Spearman rank coefficient for this correlation is r=0.84 with a a null probability of P=0.001 (Table 2).

As the HI gas that absorbs/scatters \LyA\ may have a velocity distribution similar to the HII regions in the galaxy, and increasing the velocity dispersion of the HI gas may result in larger velocity offset (e.g. Steidel et al. 2010), we normalize the blue peak velocity by the FWHM of \Ha.  
The \fesc\ shows an even tighter correlation with the ratio V(blue-peak)/FWHM(\Ha), with Spearman rank coefficient r=0.94 and null probability P=1e-5. 
Thus adding a new variable FWHM(\Ha) improves the correlation. As the FWHM(\Ha) of these Green Peas have a small range of about 180-300 \kms, we do a simple simulation to test the significance of the improvement. In each run we substitute a random number between 180-300 \kms for the FWHM(\Ha) of these Green Peas, and calculate the Spearman rank correlation between \fesc\ and V(blue-peak)/FWHM(\Ha). We run it $10^{5}$ times, and 1.8\% of the resulting correlation coefficients are better than the observed data (r=0.94). So there is 98.2\% probability that the improvement of correlation is not a random event.

The \fesc\ doesn't correlate with the red peak velocity (Spearman r=-0.2, P=0.55) or the ratio V(red-peak)/FWHM(\Ha) (Spearman r=0.04, P=0.9), while it does anti-correlate with the peak separation V(red-peak)-V(blue-peak) (Spearman r=-0.83 and P=0.002) and (V(red-peak)-V(blue-peak))/FWHM(\Ha) (Spearman r=-0.75 and P=0.007). Thus correlations of \fesc\ with velocity differences V(red-peak)-V(blue-peak) are mostly due to the stronger correlation with blue peak velocity. We also notice that in figure 2 Green Peas with large \fesc\ (GP1424+4217, GP1133+6514, GP1219+1526) show V(Valley)\textgreater0, while the others with lower \fesc\ have V(Valley)\textless 0.

In figure 5, the \fesc\ shows a very weak correlation with the flux ratio of blue and red part, Flux(blue)/Flux(red) (Spearman r=0.44, P=0.17). As the \fesc\ increases, both the red and blue peaks get stronger relative to the intrinsic \LyA\ flux. There is only a weak trend that the blue peak increases faster with \fesc\ than the red peak. In the middle panel of figure 5, the \fesc\ shows correlation with the residual flux density $f_{\lambda}(valley)$ normalized by $f_{\lambda}$(red peak) (Spearman r=0.89 and P=2e-4).
In the right panel of figure 5, \fesc\ shows anti-correlation with FWHM(red) (Spearman r=-0.83 and P=0.002). It is not clear whether \fesc\ also has weak correlation with FWHM(blue), as the blue part profiles are noisy and the uncertainties of measured FWHM(blue) are too large. The \fesc\ doesn't correlate with FWHM(\Ha) (Spearman r=0.04, P=0.9). 

Although the current sample size is small, these correlations are encouraging.  We interpret these correlations between \LyA\ escape and \LyA\ profile in section 7.1.

\begin{deluxetable}{lcl}
\tablecaption{Relations of \fesc\ and other properties}
\tablehead{\colhead{Variables} & \colhead{Spearman r} & \colhead{P} \\ 
\colhead{(1)} & \colhead{(2)} & \colhead{(3)} } 

\startdata

\fesc\ vs. FWHM(\Ha) & 0.04 & 0.9 \\
\fesc\ vs. V(red-peak)/FWHM(Ha) & 0.04 & 0.9 \\
\fesc\ vs. V(red-peak) & -0.2 & 0.55 \\
\fesc\ vs. Flux(blue)/Flux(red) & 0.44 & 0.17 \\
\fesc\ vs. E(B-V) & -0.62 & 0.05 \\
\fesc\ vs. ${V(red-peak)-V(blue-peak)}\over{FWHM(Ha)}$& -0.75 & 0.007 \\
\fesc\ vs. metallicity & -0.82 & 0.004 \\
\fesc\ vs. FWHM(red) & -0.83 & 0.002 \\
\fesc\ vs. V(red-peak)-V(blue-peak) & -0.83 & 0.002 \\
\fesc\ vs. V(blue-peak) & 0.84 & 0.001 \\
\fesc\ vs. $f_{\lambda}$(valley)/$f_{\lambda}$(red peak) & 0.89 & 0.0002 \\
\fesc\ vs. V(blue-peak)/FWHM(Ha) & 0.94 & $10^{-5}$ 
\enddata

\tablecomments{Column Descriptions: (1) Relations between \fesc\ and other properties; (2) The Spearman rank coefficient for each correlation; (3) The null probability of each correlation. }

\end{deluxetable}

\section{\LyA\ Escape, Dust and Metallicities} 
In order to probe what factors are dominant in allowing \LyA\ to escape, we also investigate relations between \fesc\ and galactic properties measured from optical spectra, namely the dust reddening E(B-V) and the metallicity. 
In the \LyA\ escape process, dust can absorb the \LyA\ photons and decrease \fesc. Multiple studies have confirmed that \LyA\ escapes more easily from galaxies with low dust reddening, and there is an anti-correlation between \fesc\ and E(B-V), although dust extinction is not the only factor to determine \fesc\ (Atek et al. 2008, 2014; Scarlata et al. 2009; Finkelstein et al. 2011; Cowie et al. 2011; Hayes et al. 2014). 
In the left panel of figure 6, we show the \fesc\ - E(B-V) relation for Green Peas. 
It is qualitatively similar to previous studies, though this Green Peas sample has higher \fesc\ and only covers a small E(B-V) range.  
There are two outliers in figure 6. One is GP0303-0759 (marked by a large circle) with the lowest metallicity and E(B-V). Its WISE three bands colors suggest it an AGN candidate (Malhotra et al. in preparation). The other one is GP1249+1234 (marked by a large square) which has single peaked line profile and large E(B-V). 
After both GP0303-0759 and GP1249+1234 are excluded, the anti-correlation between \fesc\ and E(B-V) has Spearman coefficient r=-0.62 and P=0.05 (note that r=-0.35 and P=0.28 if only the AGN candidate GP0303-0759 is excluded). It suggests dust extinction is an important factor of \LyA\ escape even in this Green Peas sample covering a small E(B-V) range. 

Dust reddening is usually higher in galaxies with higher metallicity and stellar mass. So we expect \fesc\ may also anti-correlate with metallicity. Some previous studies have found that \LyA\ escape is easier in low metallicity galaxies, while others have found no dependence on metallicity.  
Cowie et al. (2011) showed that GALEX z$\sim$0.3 LAEs have lower metallicities than UV-selected galaxies with similar UV magnitudes but no \LyA\ emission. Finkelstein et al. (2011) also showed that GALEX z$\sim$0.3 LAEs have lower metallicities than similar-mass galaxies from the SDSS.  In contrast, Atek et al. (2014) found that there is no correlation between \fesc\ and metallicity for a combined z=0 - 0.3 LAE sample. Hayes et al. (2014) showed that in the LARS sample of 14 LAEs, a few galaxies with high \fesc\ have on average lower metallicity than the other galaxies with low \fesc. 
Note that the metallicities in those studies are measured from the strong line indexes ([NII]/\Ha, ([OII]+[OIII])/\Hb, or ([OIII]/\Hb)/([NII]/\Ha) and have large uncertainties. 
For the present Green Pea sample, we show the relation between \fesc\ and metallicity measured from the more accurate $T_{e}$ method (right panel of figure 6).  If both outliers GP0303-0759 and GP1249+1234 are excluded, \fesc\ anti-correlates with metallicity with Spearman coefficient r=-0.82 and P=0.004 (note that r=-0.58 and P=0.06 if only the AGN candidate GP0303-0759 is excluded).  It is tighter than the correlation between \fesc\ and dust extinction.

\begin{figure*}[ht]
\centering
  \includegraphics[width=0.9\textwidth]{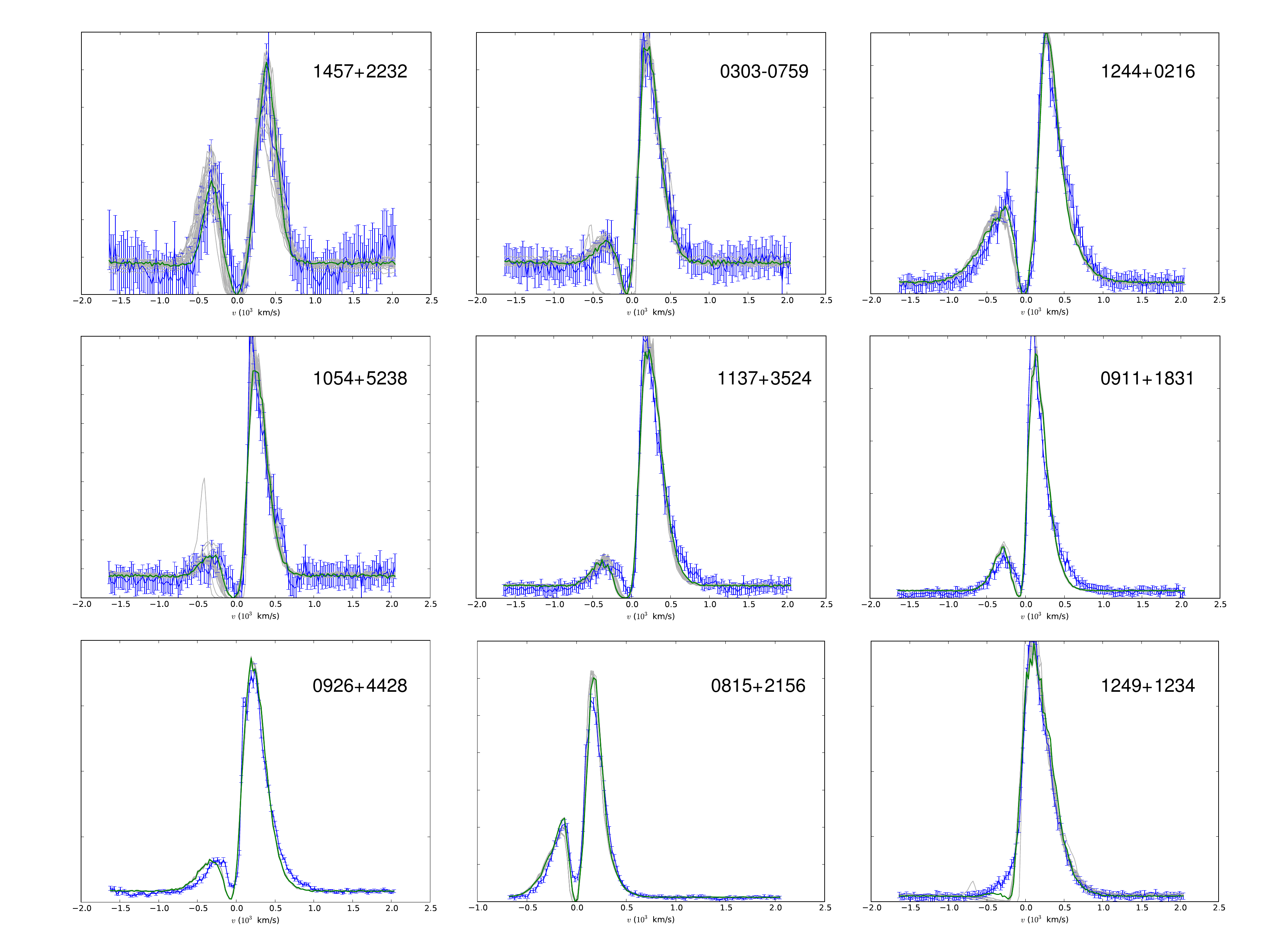}
 \caption{Comparison of the observed \LyA\ profiles (blue lines) and the best fit \LyA\ profiles (green lines) from radiative transfer models. To show the uncertainty in the fitted profiles, we also plot (grey lines) model \LyA\ profiles corresponding to 25 random samples from the MCMC chain of each galaxy. The flux of all profiles are normalized.  }
\end{figure*}

\begin{figure*}[ht]
\centering
  \includegraphics[width=0.9\textwidth]{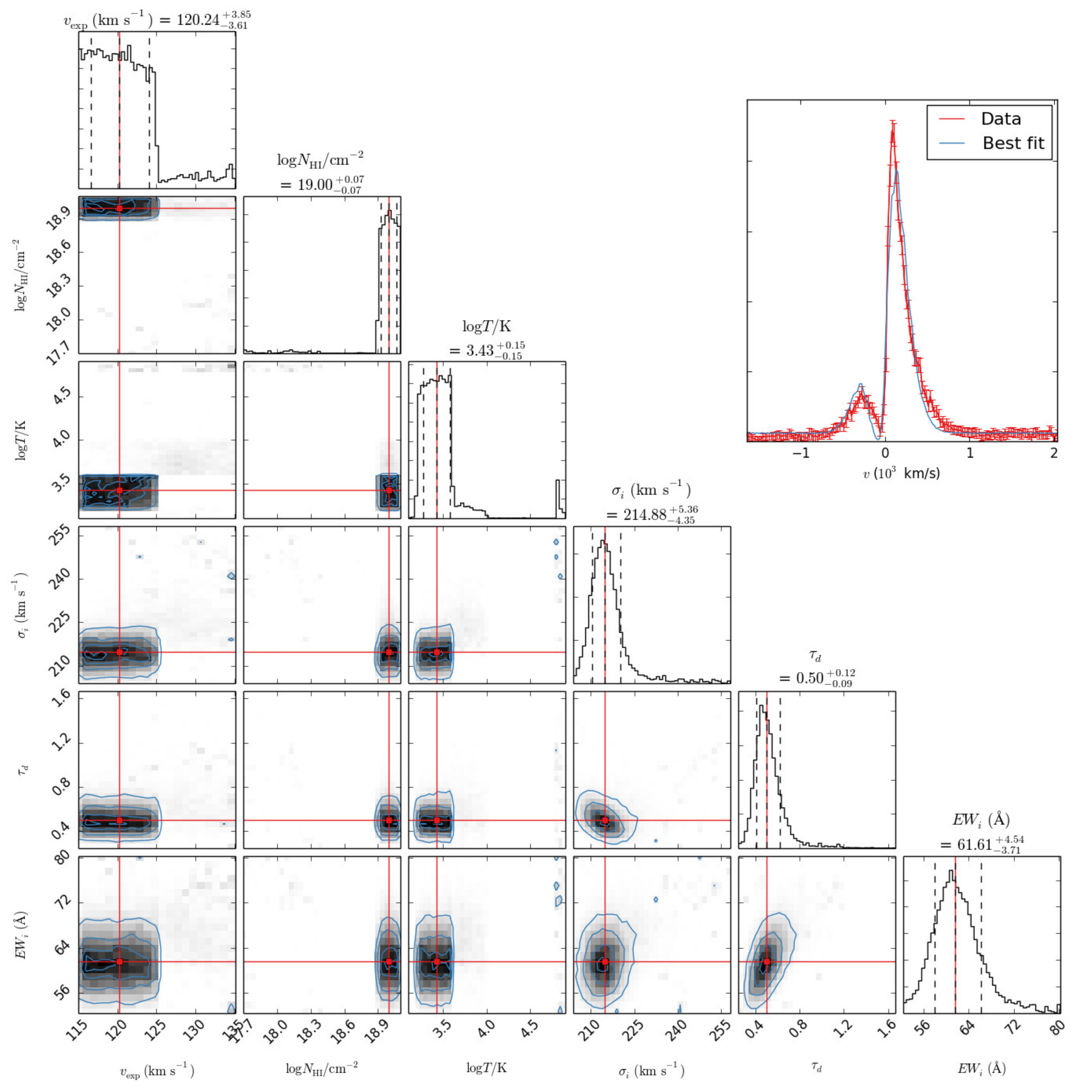}
 \caption{The fitting results for GP0911+1831. {\it Upper right}: The observed \LyA\ profile and the best fit model profile. {\it Main plot}: The one-dimensional and two dimensional marginal posterior distributions from the MCMC chains. In the plot of one-dimensional marginal posterior distribution, the dashed lines mark the 16\%, 50\%, and 84\% percentiles (numbers are shown in top of each column). In the plot of two dimensional marginal posterior distributions, the blue lines mark the (2, 1.5, 1, 0.5) $\sigma$ contours, and the gray shading gives the posterior density (Gronke et al. 2015).}
\end{figure*}

\section{\LyA\ Radiative Transfer Modeling}
To extract more information from the \LyA\ profiles and explore the physical process of \LyA\ escape, we fit the \LyA\ profiles of Green Peas with expanding HI shell radiative transfer models (Dijkstra et al. 2014; Gronke et al. 2015).  In the radiative transfer model, \LyA\ photons were generated in the shell center and then scattered/absorbed by a shell of HI gas. The intrinsic \LyA\ line has a Gaussian profile with width $\sigma$.
The shell is described by four parameters:  (a) outflow velocity $v_{exp}$, (b) hydrogen column density  $N_{HI}$, (c) temperature T (including turbulent motion as well as the true temperature), and (d) dust optical depth $\tau_{d}$. The radiative transfer simulation generates a grid of \LyA\ spectra covering large ranges of parameters. To compare the observed spectrum with models, we calculate the $\chi^{2}$ and likelihood values, and find the best matched model parameters ($\sigma$, $v_{exp}$, $N_{HI}$, T, $\tau_{d}$) using Markov Chain Monte Carlo (MCMC) methods. 
The one-dimensional and two dimensional marginal posterior distributions are calculated from the MCMC chains. The resulting parameter value and lower and upper uncertainties are the 50\%, 16\%, and 84\% percentiles respectively in the one-dimensional marginal posterior distribution of each parameter. The details of the models and fitting methods are described in Gronke et al. (2015). As an example, we show the marginal posterior distributions and fitting result for GP0911+1831 in figure 8. 

The model gives good fits for profiles of 9/12 Green Peas. For the 9 cases where the model fitting proceeded smoothly, the best-fit model profiles are shown in figure 7, and the derived parameters and uncertainties are shown in Table 4. The best-fit expanding velocities are between 27-127 \kms\ for 8 Green Peas except GP1249+1234 which has $V_{exp}$=354 \kms\ and single peak \LyA\ profile. 
The derived $N_{HI}$ are between $10^{19} - 10^{20} cm^{-2}$.

The model fails to fit the three double-peaked profiles (GP1424+4217, GP1133+6514, GP1219+1526) where the \fesc\ is high and V(Valley)\textgreater0 (see Section 4 and figure 2). 
This could be due to a few reasons: (1) ill-understood COS instrumental effects in taking spectra for slightly extended objects (as the Green Peas); (2) inflowing gas; (3) Spatial and velocity offsets of star-forming regions relative to the \LyA\ emitting gas; (4) general inapplicability of simple shell models in some cases.  We will report fitting with improved model in a following paper.  For these three cases, we manually adjust the model parameters to match the observed depth of the ``valley'' and the relative heights of the blue and red peaks. The resulting model parameters (Table 4) show outflowing velocities $\sim 20-110~$\kms\ and $N_{HI} \sim 10^{18.0}~cm^{-2}$. The resulting $N_{HI}$ is moderately reliable because a significant larger $N_{HI}$ can't fit the observed large residual flux near the line center.

We plot \fesc\ as a function of fitted $N_{HI}$ in figure~9. For the three cases that needed manual adjusting, we set generous errors of $N_{HI}$ to 0.5 dex. The results suggest an anti-correlation between \fesc\ and $N_{HI}$ with Spearman r=-0.69 and P=0.01.  Previous studies have suggested that LAEs have lower $N_{HI}$ than non-LAEs (e.g. Shibuya et al. 2014; Erb et al. 2014; Hashimoto et al. 2015). Here for the first time, we show that \fesc\ is higher at lower $N_{HI}$ in an LAE sample. Therefore we conclude that the low column density of HI gas helps \LyA\ escape in these Green Peas.

\begin{figure}[ht]
\centering
  \includegraphics[width=0.5\textwidth]{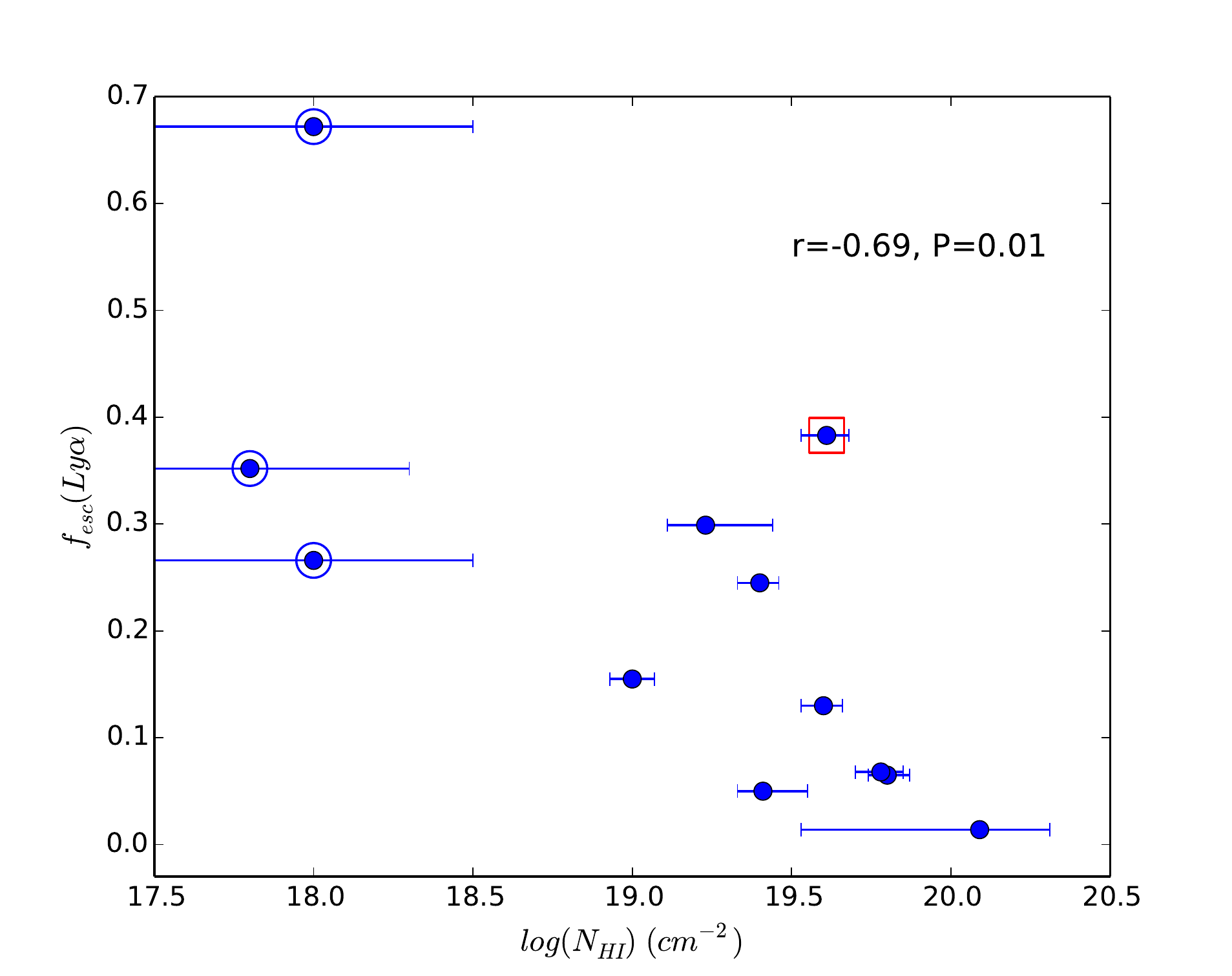}
 \caption{Relations between \fesc\ and best fit $N_{HI}$ from the expanding shell model.  $N_{HI}$ was derived by fitting of the shell model parameters for 9/12 Green Peas. For the other three cases (GP1424+4217, GP1133+6514, and GP1219+1526, marked by large circles) where the fitting procedure failed, we plot the $N_{HI}$ obtained by manually adjusting the model parameters to match the observed depth of the ``valley" and the relative heights of blue and red peaks (see Section 6). GP1249+1234 (marked by a large square) has single peak profile, large expanding velocity, and large E(B-V). The Spearman correlation coefficient r, and null probability P, are calculated with all 12 sources. }
\end{figure}

\begin{deluxetable*}{llllll}
\tablecaption{\LyA\ profile Model Parameters}
\tablehead{\colhead{ID} & \colhead{$\sigma$} & \colhead{$V_{exp}$} & \colhead{log($N_{HI}$)} & \colhead{log(T)} & \colhead{$\tau_{d}$} \\ 
\colhead{} & \colhead{(\kms)} & \colhead{(\kms)} & \colhead{($cm^{-2}$)} & \colhead{(K)} & \colhead{} \\
\colhead{(1)} & \colhead{(2)} & \colhead{(3)} & \colhead{(4)} & \colhead{(5)} & \colhead{(6)} } 
\startdata
GP1457+2232 & $280^{+53}_{-34}$ & $27^{+12}_{-12}$ & $20.09^{+0.22}_{-0.56}$ & $4.5^{+0.4}_{-1.1}$ & $0.34^{+2.11}_{-0.28}$ \\
GP0303-0759 & $245^{+17}_{-22}$ & $113^{+74}_{-14}$ & $19.41^{+0.14}_{-0.08}$ & $3.7^{+0.5}_{-0.5}$ & $0.25^{+1.98}_{-0.18}$ \\
GP1244+0216 & $351^{+17}_{-27}$ & $55^{+12}_{-7}$ & $19.80^{+0.07}_{-0.06}$ & $4.3^{+0.3}_{-0.3}$ & $0.14^{+0.25}_{-0.04}$ \\
GP1054+5238 & $226^{+12}_{-11}$ & $91^{+10}_{-5}$ & $19.78^{+0.07}_{-0.08}$ & $4.5^{+0.2}_{-0.5}$ & $0.54^{+0.33}_{-0.2}$ \\
GP1137+3524 & $239^{+7}_{-12}$ & $127^{+5}_{-23}$ & $19.60^{+0.06}_{-0.07}$ & $3.4^{+0.2}_{-0.2}$ & $0.47^{+0.14}_{-0.19}$ \\
GP0911+1831 & $215^{+5}_{-4}$ & $120^{+4}_{-4}$ & $19.00^{+0.07}_{-0.07}$ & $3.4^{+0.2}_{-0.2}$ & $0.50^{+0.12}_{-0.09}$ \\
GP0926+4428 & $266^{+3}_{-3}$ & $110^{+4}_{-3}$ & $19.40^{+0.06}_{-0.07}$ & $3.7^{+0.2}_{-0.6}$ & $0.00^{+0.01}_{-0.00}$ \\
GP0815+2156 & $207^{+6}_{-22}$ & $34^{+8}_{-13}$ & $19.23^{+0.21}_{-0.12}$ & $4.0^{+0.2}_{-0.3}$ & $0.01^{+0.25}_{-0.01}$ \\
GP1249+1234 & $206^{+10}_{-6}$ & $354^{+11}_{-7}$ & $19.61^{+0.07}_{-0.08}$ & $4.2^{+0.2}_{-1.0}$ & $1.24^{+0.34}_{-0.14}$ \\
\hline 
GP1424+4217 & 350 & 110 & 18.0 & 3.4 & 0.16 \\
GP1133+6514 & 214 & 20 & 17.8 & 3.0 & 0.16 \\
GP1219+1526 & 214 & 20 & 18.0 & 3.0 & 0.16 
\enddata
\tablecomments{Column Descriptions: (2) 1$\sigma$ width of the Gaussian profile of intrinsic \LyA\ line; (3) outflowing velocity of HI gas shell; (4) HI column density of the expanding gas shell; (5) temperature of HI gas; (6) dust optical depth. The first 9 rows show the best fit parameters and errors for nine Green Peas. The 68\% uncertainties are given for all parameters. The last 3 rows show the manually adjusting results for three profiles that the model fails to fit. }
\end{deluxetable*}

\section{Discussion}
\subsection{Correlations between \fesc\ and \LyA\ Profiles}
High quality \LyA\ profiles of Green Peas provide a good opportunity to study \LyA\ escape in LAEs. 
We show that the \fesc\ is correlated with kinematic features of \LyA\ escape. 
What are the implications of these correlations about \LyA\ escape? 

\textbf{(1) \fesc\ and V(blue-peak)}: As shown in Section 4, the \fesc\ correlates strongly with V(blue-peak) and peak separation. Both V(red-peak) and V(blue-peak) have a complex dependence on $N_{HI}$, gas outflows, and gas temperature. 
The velocity offset V(blue-peak) is generally larger when there is more HI gas and a larger numbers of scatterings for \LyA\ photons (e.g. Verhamme et al. 2015). So the correlation between \fesc\ and V(blue-peak) suggests LAEs with higher \fesc\ have less scatterings for \LyA\ photons and lower $N_{HI}$.

\textbf{(2) \fesc\ and $f_{\lambda}$(valley)/$f_{\lambda}$(red-peak)}: A few Green Peas show significant residual flux near \LyA\ line center. As the optical depth of HI gas at the systemic redshift increases rapidly with $N_{HI}$, only the Green Peas with low $N_{HI}$ will have residual flux at \LyA\ line center.
Low $N_{HI}$ should correspond to high escape fraction, and indeed
we see that  \fesc\ is correlated with
$f_{\lambda}$(valley)/$f_{\lambda}$(red-peak).

\textbf{(3) \fesc\ and FWHM(red)}: \fesc\ is anti-correlated with FWHM(red). To broaden the red peak profile generally requires scattering of \LyA\ photons to redder frequency by HI gas. Again, more scattering results in more dust extinction and smaller \fesc.

\subsection{Difficulties of Expanding HI Shell Approximations}

We fit the \LyA\ profiles with an expanding single shell radiative transfer model. 
While the models often yielded remarkably good fits to the observed line profiles,
they have limitations that we discuss here.

\textbf{Velocity of central valley of \LyA\ profile:} The single shell radiative transfer model can't fit the three profiles with $V (valley) > 0$.   Simply allowing a negative shell velocity (corresponding to a single infalling shell) would permit $V(valley)>0$, but is not a solution, because it also produces profiles whose blue peak is higher than the red peak (contrary to observation). 
One possibility is that besides the expanding HI shell component, there is another inflowing HI component that absorbs more \LyA\ photons in the red side of profile and make $V (valley) > 0$. 
Another possibility is that there are a few star-forming gas clumps with different central velocities. 
Thus there could be spatial and velocity offsets of star-forming regions relative to the \LyA\ emitting gas. 
If the \LyA\ emissions are dominated by the gas clumps with redshifted velocities relative to the average systemic redshift, then the observed total \LyA\ profile would show a red-ward offset and $V (valley) > 0$. 
Generally, to fit these profiles requires us to consider more realistic HI gas distributions than the HI shell approximation.

\textbf{\LyA\ intrinsic line width}:
The model fitting results show intrinsic \LyA\ line width between 200 - 350 \kms\ (1 $\sigma$ of Gaussian profile) and FWHM about 470-820 \kms, which are  $\sim$ 2-3 times larger than the intrinsic \Ha\ line width. If the intrinsic \LyA\ profile is as narrow as \Ha\ profile, producing the observed broad \LyA\ profiles would require large $N_{HI}$ and a high HI gas temperature $T$. However larger $N_{HI}$ and T will also increase absorption of \LyA\ photons, and result in strong central absorption and weak blue peak, which contradict the observed profile. Hashimoto et al. (2015) fit \LyA\ profiles of a few z$\sim$2 LAEs and also found the best fit intrinsic \LyA\ width is a few times larger than the intrinsic \Ha\ or \oiii width. \LyA\ and \Ha\ powered by star formation ought to originate in the same gas, so a wider intrinsic \LyA\ profile would suggest either that not all the \LyA\ is powered by star formation (cf. Hashimoto et al. 2015), or that there are important radiative transfer effects that broaden \LyA\ nearer to the source, before it encounters the model's expanding outer HI shell.

\textbf{Shell expansion velocity}:
The resulting shell expansion velocities are between 20-130 \kms\ for 11/12 Green Peas (and larger for the single-peaked line in GP1249+1234). These expansion velocities are comparable to the 1$\sigma$ width of \Ha\ emission lines (80-130 \kms), but are about 2 times smaller than the average outflow velocities traced by low ionization metal absorption lines ($\sim$ 80-300 \kms, see Henry et al. 2015). This may suggest that the low ionization metal absorption lines trace a different outflowing gas component from the HI gas, and that the velocity distribution of HI gas is more similar to that of HII regions.

\subsection{Metallicity and \LyA\ Escape}

We showed that \fesc\ anti-correlates with metallicity. The \fesc\ is also higher in Green Peas with lower stellar mass and higher sSFR. However, only dust and HI gas directly interact with \LyA\ photons. What causes the correlation of \fesc\ and metallicity? It is not simply caused by a correlation between the abundances of metal and dust, as the metallicity has tighter correlation with \fesc\ than dust reddening.  Thus the \fesc - metallicity correlation may suggest lower $N_{HI}$ in lower metallicity Green Peas.  As the mass and metallicity are correlated, one possibility is galaxies with lower mass and lower metallicity have less HI gas.  The other possibility is outflows in lower metallicity galaxies blow out more holes with low $N_{HI}$ and increase \fesc.

Outflows have a large impact on the metallicity of low mass galaxies. These emission line selected galaxies (including LAEs and Green Peas) can have metallicities as low as 1 dex below the usual mass - metallicity relation (e.g.  Xia et al. 2012; Ly et al. 2014; Song et al. 2014). The galactic mass - metallicity relation is generally a result of interactions between the metal enrichment and gas outflows: star formation enriches the interstellar medium, but outflows can drive metal enriched gas out and decrease the metallicity.  Since outflows are more effective where the potential is shallow, this process results in lower metallicities at lower masses. Emission line galaxies like LAEs and Green Peas tend to have compact star formation and high sSFR,  resulting in stronger outflows and lower metallicity than other galaxies with similar mass. The outflows from stellar feedback are dominated by ionized gas. These ionized outflows can drive HI gas out or make ionized holes in interstellar medium, thus reducing the effective column of $N_{HI}$ for \LyA\ photons.

\section{Conclusion}

We have investigated high quality \LyA\ spectra of a sample of 12 Green Peas.  {\it All} show \LyA\ emission lines. We compared the EW(\LyA) distribution of this Green Peas sample to high redshift LAEs. From the \Ha\ emission lines we obtained the systemic redshift and intrinsic \LyA\ emission. The high S/N \LyA\ spectra permit us to measure accurately the blue peaks of \LyA. So we measured the velocities and width of the \LyA\ profiles and explored correlations of profile features with \LyA\ escape fraction. 
We also compared \fesc\ with dust reddening and metallicity. In addition, we fit \LyA\ profiles with radiative transfer model and discussed constraints on HI gas and \LyA\ escape. Our main results are as follows.

\begin{enumerate}

\item The EW(\LyA) distribution of this Green Pea sample is very similar to our z=2.8 LAE sample.  As these Green Peas have small mass, compact size, and high emission line equivalent widths, similar to high-z LAEs, we conclude that Green Peas are the best analogs of high-z LAEs in the local universe.

\item The \fesc\ shows correlations with the blue peak velocity of \LyA, the ratio V(blue-peak)/FWHM(\Ha), the flux density ratio $f_{\lambda}$(valley)/$f_{\lambda}$(red-peak), and the width of \LyA\ red peak FWHM(red). As more scatterings in HI gas can make the \LyA\ blue peak bluer, the residual central flux smaller, and the red peak profile wider, these correlations strongly suggest low $N_{HI}$ and fewer scatterings help the \LyA\ photons escape.

\item The single shell radiative transfer models can reproduce most profiles, and get column densities of HI gas. The resulting $N_{HI}$ anti-correlates with \fesc\ in this LAE sample, again indicating that 
low $N_{HI}$ is key to \LyA\ escape.

\item However, the single shell model also has difficulties. It fails to fit three profiles. 
The best-fit intrinsic line width is 2-3 times larger than the \Ha\ width. The best-fit HI gas outflow velocity is small and only about half of the average velocity of low ionization absorption lines.  Fully reproducing all
features of the observed lines will require modeling \LyA\ radiative transfer in more realistic HI gas distributions.

\item In this Green Pea sample, \fesc\ shows a weak anti-correlation with dust reddening, and a  stronger anti-correlation with metallicity.  We suggest that this correlation may be an effect of the mass-metallicity relation and trends of other properties with galaxy mass. Lower metallicity galaxies are likely to have less HI gas and dust. In addition, ionized gas outflows can blow out the metal enriched gas, making holes with low $N_{HI}$ and helping \LyA\ escape.

\end{enumerate}

In conclusion, Green Peas provide an unmatched opportunity to study \LyA\ escape in LAEs. 
Our results suggest that LAEs with high \LyA\ escape fraction have low metallicity, low HI column density, and mild HI gas outflow. 
In future work, we will compare \LyA\ profiles and the correlations found in this work with improved models and provide more quantitative constraints on the HI gas and \LyA\ escape in LAEs.

\acknowledgments
We thank M. S. Oey for helpful discussions.  H. Y. acknowledges support from China Scholarship Council.


\begin{thebibliography}{}

\bibitem[Ahn et al.(2001)]{2001ApJ...554..604A} Ahn, S.-H., Lee, H.-W., 
\& Lee, H.~M.\ 2001, \apj, 554, 604 

\bibitem[Amor{\'{\i}}n et al.(2014)]{2014ApJ...788L...4A} Amor{\'{\i}}n, 
R., Grazian, A., Castellano, M., et al.\ 2014, \apjl, 788, L4 

\bibitem[Atek et 
al.(2008)]{2008A&A...488..491A} Atek, H., Kunth, D., Hayes, M., {\"O}stlin, G., \& Mas-Hesse, J.~M.\ 2008, \aap, 488, 491 

\bibitem[Atek et 
al.(2009)]{2009A&A...502..791A} Atek, H., Schaerer, D., \& Kunth, D.\ 2009, \aap, 502, 791 

\bibitem[Atek et 
al.(2014)]{2014A&A...561A..89A} Atek, H., Kunth, D., Schaerer, D., et al.\ 2014, \aap, 561, A89 

\bibitem[Bond et al.(2010)]{2010ApJ...716L.200B} Bond, N.~A., Feldmeier, 
J.~J., Matkovi{\'c}, A., et al.\ 2010, \apjl, 716, L200 

\bibitem[Calzetti et al.(2000)]{Calzetti} 
Calzetti, D., Armus, L., Bohlin, R.~C., et al. 2000, \apj, 533, 682 

\bibitem[Cardamone et al.(2009)]{Cardamone} 
Cardamone, C., Schawinski, K., Sarzi, M., et al. 2009, \mnras, 399, 1191  

\bibitem[Charlot \& Fall(1993)]{1993ApJ...415..580C} Charlot, S., \& Fall, S.~M.\ 1993, \apj, 415, 580 

\bibitem[Chonis et al.(2013)]{2013ApJ...775...99C} Chonis, T.~S., Blanc, 
G.~A., Hill, G.~J., et al.\ 2013, \apj, 775, 99

\bibitem[Cl{\'e}ment et al.(2012)]{2012A&A...538A..66C} Cl{\'e}ment, B., Cuby, J.-G., Courbin, F., et al.\ 2012, \aap, 538, A66 

\bibitem[Cowie et al.(2011)]{Cowie11}
Cowie, L.~L., Barger, A.~J., \& Hu, E.~M. 2011, \apj, 238, 136

\bibitem[Deharveng et al.(2008)]{Deharveng} 
Deharveng, J.-M.,  Small, T.,  Barlow, T.~A.,  et al. 2008, \apj, 680, 1072 

\bibitem[Dey et al.(1998)]{1998ApJ...498L..93D} Dey, A., Spinrad, H., 
Stern, D., Graham, J.~R., \& Chaffee, F.~H.\ 1998, \apjl, 498, L93 

\bibitem[Dijkstra et al.(2006)]{2006ApJ...649...14D} Dijkstra, M., Haiman, 
Z., \& Spaans, M.\ 2006, \apj, 649, 14 

%\bibitem[Dijkstra(2014)]{2014PASA...31...40D} Dijkstra, M.\ 2014, \pasa, 31, e040
\bibitem[Dijkstra(2014)]{2014arXiv150403693} Dijkstra, M.\ 2014, arXiv:1504.03693

\bibitem[Erb et al.(2014)]{Erb14} 
Erb, D.~K., Steidel, C.~C., Trainor, R., et al. 2014 \apj, 795, 33

\bibitem[Finkelstein et al.(2008)]{2008ApJ...678..655F} Finkelstein, S.~L., 
Rhoads, J.~E., Malhotra, S., Grogin, N., \& Wang, J.\ 2008, \apj, 678, 655 

\bibitem[Finkelstein et al.(2009)]{2009ApJ...703L.162F} Finkelstein, S.~L., 
Cohen, S.~H., Malhotra, S., et al.\ 2009, \apjl, 703, L162 

\bibitem[Finkelstein et al.(2011)]{2011ApJ...733..117F} Finkelstein, S.~L., 
Cohen, S.~H., Moustakas, J., et al.\ 2011, \apj, 733, 117 

\bibitem[Fitzpatrick(1999)]{Fitzpatrick} 
Fitzpatrick, E.~L. 1999, \pasp, 111, 63 

\bibitem[France et al.(2009)]{2009ApJ...707L..27F} France, K., Beasley, M., Keeney, B.~A., et al.\ 2009, \apjl, 707, L27 

\bibitem[Gawiser et al.(2006)]{2006ApJ...642L..13G} 
Gawiser, E., van Dokkum, P.~G., Gronwall, C., et al.\ 2006, \apjl, 642, L13 

\bibitem[Gawiser et al.(2007)]{2007ApJ...671..278G} Gawiser, E., Francke, 
H., Lai, K., et al.\ 2007, \apj, 671, 278 

\bibitem[Giavalisco et al.(1996)]{Giavalisco}
Giavalisco, M., Koratkar, A., \& Calzetti, D. 1996, \apj,  466, 831

\bibitem[Gronke et al.(2015)]{2015ApJ...812..123G} Gronke, M., Bull, P., 
\& Dijkstra, M.\ 2015, \apj, 812, 123 

\bibitem[Guaita et al.(2010)]{2010ApJ...714..255G}
 Guaita, L., Gawiser, E., Padilla, N., et al.\ 2010, \apj, 714, 255

\bibitem[Hashimoto et al.(2013)]{2013ApJ...765...70H} Hashimoto, T., Ouchi, 
M., Shimasaku, K., et al.\ 2013, \apj, 765, 70

\bibitem[Hashimoto et al.(2015)]{2015ApJ...812..157H} Hashimoto, T., 
Verhamme, A., Ouchi, M., et al.\ 2015, \apj, 812, 157 


\bibitem[Hayes et 
al.(2005)]{2005A&A...438...71H} Hayes, M., {\"O}stlin, G., Mas-Hesse, J.~M., et al.\ 2005, \aap, 438, 71 

\bibitem[Hayes et al.(2014)]{Hayes14} 
Hayes, M., \"Ostlin, G., Duval, F., et al. 2014,   \apj, 782, 6 

\bibitem[Heckman et al.(2011)]{Heckman11} 
Heckman, T.~M., Borthakur, S., Overzier, R., et al. 2011, \apj, 730, 5 

\bibitem[Henry et al.(2015)]{2015ApJ...809...19H} Henry, A., Scarlata, C., 
Martin, C.~L., \& Erb, D.\ 2015, \apj, 809, 19 


\bibitem[Hu et al.(1998)]{1998ApJ...502L..99H} Hu, E.~M., Cowie, L.~L., 
\& McMahon, R.~G.\ 1998, \apjl, 502, L99 

\bibitem[Hu et al.(2010)]{2010ApJ...725..394H} Hu, E.~M., Cowie, L.~L., Barger, A.~J., et al.\ 2010, \apj, 725, 394 

\bibitem[Izotov et al.(2011)]{Izotov11} 
Izotov, Y.~I., Guseva, N.~G., \& Thuan, T. 2011, \apj, 728, 161

\bibitem[James et al.(2014)]{2014ApJ...795..109J} James, B.~L., Aloisi, A., 
Heckman, T., Sohn, S.~T., \& Wolfe, M.~A.\ 2014, \apj, 795, 109 

\bibitem[Jaskot \& Oey(2014)]{Jaskot14} 
Jaskot, A.~E. \& Oey, M.~S. 2014, \apj, 791, 19L

\bibitem[Kashikawa et al.(2011)]{2011ApJ...734..119K} Kashikawa, N., Shimasaku, K., Matsuda, Y., et al.\ 2011, \apj, 734, 119 

\bibitem[Kunth et al.(1998)]{Kunth98}
Kunth, D., Mas-Hess, J. M., Terlevich, E., et al. 1998, \aap, 334, 11 

\bibitem[Leitherer et al.(2011)]{2011AJ....141...37L} Leitherer, C., 
Tremonti, C.~A., Heckman, T.~M., \& Calzetti, D.\ 2011, \aj, 141, 37

\bibitem[Ly et al.(2014)]{2014ApJ...780..122L} Ly, C., Malkan, M.~A., 
Nagao, T., et al.\ 2014, \apj, 780, 122 

\bibitem[Malhotra 
\& Rhoads(2004)]{2004ApJ...617L...5M} Malhotra, S., \& Rhoads, J.~E.\ 2004, \apjl, 617, L5 

\bibitem[Malhotra et al.(2012)]{2012ApJ...750L..36M} Malhotra, S., Rhoads, J.~E., Finkelstein, S.~L., et al.\ 2012, \apjl, 750, L36 

\bibitem[Mas-Hesse et al.(2003)]{2003ApJ...598..858M} Mas-Hesse, J.~M., Kunth, D., Tenorio-Tagle, G., et al.\ 2003, \apj, 598, 858 

\bibitem[Matthee et al.(2014)]{2014MNRAS.440.2375M} Matthee, J.~J.~A., 
Sobral, D., Swinbank, A.~M., et al.\ 2014, \mnras, 440, 2375 

\bibitem[McLinden et al.(2011)]{2011ApJ...730..136M} McLinden, E.~M., 
Finkelstein, S.~L., Rhoads, J.~E., et al.\ 2011, \apj, 730, 136 

\bibitem[McLinden et al.(2014)]{2014MNRAS.439..446M} McLinden, E.~M., 
Rhoads, J.~E., Malhotra, S., et al.\ 2014, \mnras, 439, 446 

\bibitem[Nakajima et al.(2012)]{2012ApJ...745...12N} Nakajima, K., Ouchi, 
M., Shimasaku, K., et al.\ 2012, \apj, 745, 12 


\bibitem[Neufeld (1990)]{1990ApJ...350..216N} Neufeld, D. A. 1990, \apj\ 350, 216

\bibitem[{\"O}stlin et al.(2009)]{2009AJ....138..923O} {\"O}stlin, G., 
Hayes, M., Kunth, D., et al.\ 2009, \aj, 138, 923 

\bibitem[{\"O}stlin et al.(2014)]{2014ApJ...797...11O} {\"O}stlin, G., 
Hayes, M., Duval, F., et al.\ 2014, \apj, 797, 11 

\bibitem[Ota et al.(2010)]{Ota10}
Ota, J., Iye, M., Kashikawa, N., et al. 2010, \apj, 722, 803


\bibitem[Ouchi et al.(2003)]{2003ApJ...582...60O} Ouchi, M., Shimasaku, K., 
Furusawa, H., et al.\ 2003, \apj, 582, 60 

\bibitem[Ouchi et al.(2010)]{2010ApJ...723..869O} Ouchi, M., Shimasaku, K., Furusawa, H., et al.\ 2010, \apj, 723, 869 


\bibitem[Pardy et al.(2014)]{2014ApJ...794..101P} Pardy, S.~A., Cannon, 
J.~M., {\"O}stlin, G., et al.\ 2014, \apj, 794, 101 


\bibitem[Pentericci et al.(2014)]{2014ApJ...793..113P} Pentericci, L., 
Vanzella, E., Fontana, A., et al.\ 2014, \apj, 793, 113 

\bibitem[Pirzkal et al.(2007)]{2007ApJ...667...49P} Pirzkal, N., Malhotra, 
S., Rhoads, J.~E., \& Xu, C.\ 2007, \apj, 667, 49 


\bibitem[Rhoads et al.(2000)]{2000ApJ...545L..85R} Rhoads, J.~E., Malhotra, S., 
Dey, A., et al.\ 2000, \apjl, 545, L85 

\bibitem[Rhoads et al.(2003)]{2003AJ....125.1006R} Rhoads, J.~E., Dey, A., 
Malhotra, S., et al.\ 2003, \aj, 125, 1006 


\bibitem[Rivera-Thorsen et al.(2015)]{2015ApJ...805...14R} Rivera-Thorsen, 
T.~E., Hayes, M., {\"O}stlin, G., et al.\ 2015, \apj, 805, 14 


\bibitem[Scarlata et al.(2009)]{Scarlata09} 
Scarlata, C., Colbert, J., Teplitz, H.~I., et al. 2009, \apj, 705, 98L


\bibitem[Schaerer et 
al.(2011)]{2011A&A...531A..12S} Schaerer, D., Hayes, M., Verhamme, A., \& Teyssier, R.\ 2011, \aap, 531, A12 


\bibitem[Schlafly \& Finkbeiner(2011)]{Schlafly} 
Schlafly, E.~F. \& Finkbeiner, D.~F. 2011, \apj, 737, 103 

\bibitem[Shapley et al.(2003)]{Shapley03} 
Shapley, A.~E., Steidel, C.~C., Pettini, M., \& Adelberger, K.~L. 2003, \apj, 588, 65 

\bibitem[Shibuya et al.(2012)]{2012ApJ...752..114S} Shibuya, T., Kashikawa, 
N., Ota, K., et al.\ 2012, \apj, 752, 114 

\bibitem[Shibuya et al.(2014)]{Shibuya} 
Shibuya, T., Ouchi, M., Nakajima, K., et al. 2014, \apj, 788, 74

\bibitem[Song et al.(2014)]{2014ApJ...791....3S} Song, M., Finkelstein, 
S.~L., Gebhardt, K., et al.\ 2014, \apj, 791, 3 

\bibitem[Stark et al.(2011)]{2011ApJ...728L...2S} Stark, D.~P., Ellis, 
R.~S., \& Ouchi, M.\ 2011, \apjl, 728, L2 


\bibitem[Steidel et al.(2010)]{Steidel10} 
Steidel, C.~C., Erb, D.~K., Shapley, A.~E., et al. 2010, \apj, 717, 289  


\bibitem[Tilvi et al.(2014)]{2014ApJ...794....5T} Tilvi, V., Papovich, C., 
Finkelstein, S.~L., et al.\ 2014, \apj, 794, 5 

\bibitem[Treu et al.(2012)]{Treu12} 
Treu, T., Trenti, M., Stiavelli, M., Auger, M.~W., \& Bradley, L.~D. 2012, \apj, 747, 27 

\bibitem[Verhamme et al.(2006)]{2006A&A...460..397V} Verhamme, A., Schaerer, D., \& Maselli, A.\ 2006, \aap, 460, 397 

\bibitem[Verhamme et 
al.(2015)]{2015A&A...578A...7V} Verhamme, A., Orlitov{\'a}, I., Schaerer, D., \& Hayes, M.\ 2015, \aap, 578, A7 


\bibitem[Wofford et al.(2013)]{Wofford} 
Wofford, A., Leitherer, C., \& Salzer, J. 2013, \apj, 765, 118 

\bibitem[Xia et al.(2012)]{2012AJ....144...28X} Xia, L., Malhotra, S., 
Rhoads, J., et al.\ 2012, \aj, 144, 28 

\bibitem[Zheng et al.(2012)]{2012ApJ...746...28Z} Zheng, Z.-Y., Malhotra, 
S., Wang, J.-X., et al.\ 2012, \apj, 746, 28 

\bibitem[Zheng et al.(2013)]{2013MNRAS.431.3589Z} Zheng, Z.-Y., 
Finkelstein, S.~L., Finkelstein, K., et al.\ 2013, \mnras, 431, 3589 


\end{thebibliography}
\end{document}